\documentclass[%
 reprint,
 amsmath,amssymb,
 aps,
floatfix,
]{revtex4-1}
\usepackage[utf8]{inputenc}
\usepackage{braket}
\usepackage{amsmath}
\usepackage{ mathrsfs }
\usepackage{verbatim}
\usepackage{graphicx}
\usepackage{array}

\begin{document}

\title[Article Title]{A Method to Detect Quantum Coherent Transport in Memristive Devices.}

\author{C Huggins}
\author{S Savel’ev}
\author{A Balanov}
\author{A Zagoskin}

\affiliation{Physics Department, Loughborough University, Loughborough, United Kingdom}

\begin{abstract}
    While the size of functional elements in memristors becomes of the orders of nano-meters or even smaller, the quantum effects in their dynamics can significantly influence their transport properties, consistent with recent experimental observations of conductance quantisation in memristors. This requires the development of experimentally realizable procedures to detect quantumness of memristors. Here we developed an experimental protocol allowing us to find evidence that the memristor can be in a superposition of states with different memristivities. We simulate the nonlinearity induced in the quantum memristive system via periodic projective measurements, observing how it manifests itself in the emergence of additional spectral components in the response to the harmonic signal. Moreover, the response demonstrates a resonant behaviour when the frequency of the projective measurements commensurates with the frequency of the input. We demonstrate that observation of such harmonic mixing can be used as experimental evidence of quantum effects in memristors.
    \end{abstract}


\maketitle

\section{Introduction}\label{sec1}
Memristors were first proposed back in 1971 \cite{Chua1971} as a logically necessary complement to the fundamental lumped circuit elements (resistors, capacitors, and inductors). These elements parameterise the relations between the dynamical variables of the circuit: current \textit{I}, charge \textit{q}, voltage \textit{V} and magnetic flux $\varphi$. Memristors were the “missing link” (directly relating $\varphi$ to \textit{q}); its appearance restores the symmetry between two active (resistor and memristor) and two reactive (capacitor and inductor) circuit elements. The term (a portmanteau of “memory” and “resistor”) reflects that a memristor can be considered a resistor, the resistance of which depends on the cumulative \textit{q} or $\varphi$ (i.e., the integral over time of \textit{I} or \textit{V} on the device) \cite{Chua2019}.
\par
It is well known that reactive circuit elements have their fully quantum analogues (see, e.g., \cite{zago11}) and can be realised via qubit-based structures, which demonstrate quantum superpositions of states with different inductances (capacitances). For active circuit elements, such behaviour may seem impossible due to the inevitable dissipation, detrimental to quantum coherence. Nevertheless, such a conclusion would be hasty. It is well established, for example, that in many mesoscopic structures the region which determines the resistance is spatially separate from those regions where the dissipation and relaxation take place \cite{Yi2016,zago11}. This provides a time scale on which a quantum superposition of states with different resistances can exist. The same possibility cannot be denied to memristances.
\par
In this paper we investigate experimentally accessible signatures of such paradoxical behaviour, to wit, the nonlinear effects arising from periodic measurements of such a quantum memristor. We look at the effects of regular measurement on spectral properties of the proposed device.

\section{Classical Memristor}\label{sec2}
First, we consider the general case of a classical current-driven memristance \textit{M} at a time \textit{t}. The memristance depends on the state variables \textit{x}, where \textit{M} and \textit{x} are described by the following: \cite{Chua2019}
\begin{align}
    V = & M(x,I,t) I \label{eq:ohm}\\
    \frac{dx}{dt}= & f(x,I,t) \label{eq:dynamics}
\end{align}
\par
\begin{figure}[h]
    \centering
    \includegraphics[width=\linewidth]{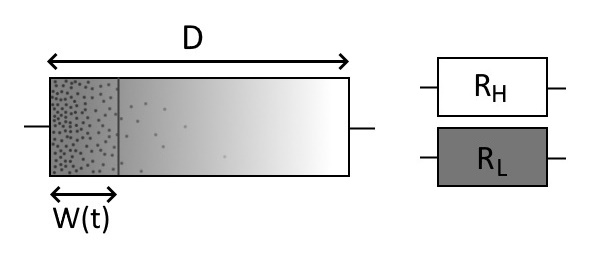}
    \caption{Diagram of an example diffusive memristor implemented via a thin film semiconductor. The width of the doped region W(t) across the active region of the device (total width D) varies with the bias applied to the film. The fully undoped state corresponds to the highest memrisistance, $M(X=0)=R_H$, and the fully doped corresponds to the lowest, $M(X=1)=R_L$.}
    \label{fig:linearion}
\end{figure}

For our purposes, we utilise a memristive function based on the well-studied diffusion memristors \cite{Yi2016,Found2008,Wang2017,Prezioso2016,Jiang2017,USHAKOV2021110803,USHAKOV2021110997,https://doi.org/10.1002/adfm.201600680}. Though our scheme can be generalised to any other mechanism for memristive switching, e.g., those outlined in ref \cite{Sun2019}. The state variable for our diffusion memristor is the (normalised) width of the doped region, \textit{X}(t)=\textit{W}(t)/\textit{D} (see Fig \ref{fig:linearion}) where \textit{W}(t) is the width of the doped region, and D is the total width of the active part of the device (\textit{W} is assumed to be constant in time). The fewer the charge carriers, the larger the width of the undoped region, corresponding to a higher memristance. A sinusoidal input current will drive the memristance to oscillate between minimum/maximum memristance values, $R_L$ and $R_H$, where $\mu_v$ is the average ion mobility\cite{Found2008}: 
\begin{equation}
     M(q)= R_H ( 1-\frac{\mu_v}{D^2} R_L q) \label{eq:linearq}
\end{equation}
\par
The voltage response (for a sinusoidal input current $I=I_0\sin(\omega_st)$, with the initial condition q(t=0)=0) is expected to have two dominant frequencies. One at the source frequency $\omega_s$, and one at the second harmonic, $2\omega_s$. This second order response arises from the multiplication of q and I.
\begin{equation}
\begin{split}
    V_C =  \left( R_HI_0-\frac{\mu_v}{D^2}R_{L}R_{H}\frac{I_0^2}{\omega_s} \right) \sin(\omega_st) +...\\ +\frac{\mu_v}{2D^2}R_{L}R_{H}\frac{I_0^2}{\omega_s} \sin(2\omega_st) \label{eq:classical}
\end{split}
\end{equation}
\par 
All classical, periodically driven memristors will have a response at $\omega_s$. In the next sections, we look for differences in the spectra between the classical and quantum cases.

\section{Quantum Model}\label{sec3}
We model a qubit-controlled quantum memristor. Experimentally, this could describe a setup with a van der Waals heterostructure \cite{VDW} or a singularly charged pair of quantum dots, where electrons in each state get transported to different memristive filaments. Measuring the qubit gives an eigenvalue \textit{j}, determining the memristance with \textit{j}=0,1, corresponding to qubit states $\ket{0}$, $\ket{1}$.
\par
\begin{figure}[h]
    \centering
    \includegraphics[width=\linewidth]{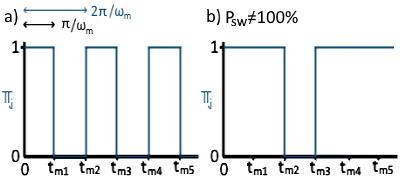}
    \caption{Square wave switching functions $\Pi_j$. a) is an example $\Pi_0$, with a 100\% switching probability, $P_sw=1$. The arrows demonstrate the period of the square wave, $2\pi/\omega_m$. b) is an example with a lower $P_{sw}$=0.4. }
    \label{fig:squarewaves}
\end{figure}

The state evolves via the Schr\"odinger equation, determining probability of switching state at each measurement ($P_{sw}$). We describe state switching via pulse functions $\Pi_j(t)$. If the state is measured \textit{j} on the $k^{th}$ measurement, then over the proceeding time interval (between times
[$t_{mk}$,$t_{mk+1}$]), $\Pi_j=1$ and $\Pi_{1-j}=0$. This acts as an on/off switch for each state, illustrated in fig. \ref{fig:squarewaves}. 
\begin{equation}
    V_Q(t)=\sum_{j=0}^1 \Pi_j(t)  V_{Cj}(t) \label{eq:vq}
\end{equation}
\par
Pulse widths in $\Pi_j$ are characterised by the statistics of measurement events and $P_{sw}$. $P_{sw}$ acts as a mediator between the classical and quantum cases. The $P_{sw}=0$ case corresponds to a classical system of electron transport through only one of the memristive filaments (no superpositions of electron position). $P_{sw}$ depends on the state, governed by the Schr\"odinger equation. We utilise a pseudospin Hamiltonian for the purposes of this report (equation (\ref{eq:Hmeas})) where $\Delta$ and $\epsilon$ have been chosen to give a high probability of switching: 
\begin{gather}  
 H_0=-\frac{1}{2}(\Delta \sigma_x+\epsilon \sigma_z)\label{eq:Hmeas}\\
    \frac{d}{dt}\ket{\psi}=\frac{1}{i\hbar}H_0\ket{\psi} \label{eq:schro}
\end{gather}
\par 
We also consider examples where there is some uncertainty in the measurement frequency (modelling realistically imperfect measurement protocols). This uncertainty is normally distributed around the average measurement frequency $\omega_m$, by a variance of $\sigma_\omega^2$.
\par
We predict the output spectrum using equation (\ref{eq:vq}), $\tilde{V}_Q(\omega)$ is the convolution of $\tilde{\Pi}_j(\omega)$ and $\tilde{V}_{Cj}(\omega)$ (equation (\ref{eq:conv})). To determine $\tilde{\Pi}_j(\omega)$, we use the ideal model for $\Pi_j(t)$, a periodic square wave of frequency $\omega_m/2\pi$ (halved since 4 measurements are needed for a full wave form, see fig. \ref{fig:squarewaves}). This model assumes a 100\% switching probability, a suitable approximation, since we selected a Hamiltonian to give a high switching rate (at our measurement frequency).
\par
The most significant peaks in $\tilde{\Pi}_j(\omega)$ are at odd multiples of this square-wave frequency $\omega_m/2\pi$. To find $\tilde{\Pi}_j(\omega)$ (equation (\ref{eq:PI})) we first found the Fourier coefficients for $\Pi_j(t)$, before using this to calculate the one-sided Fourier transform. Considering $V_Q$ we assume that $\overline{\psi_j}$ is the time-average of $\Pi_j$ (for a high switching rate, this will be $\approx$0.5). We also consider that the time-average of  $V_{Cj}$ is negligible. For brevity, let $A_{2\omega_s}=\frac{\mu_v}{2D^2}R_{Lj}R_{Hj}\frac{I_0^2}{\omega_s}$:
\begin{align}
    \tilde{\Pi}_j(\omega) \approx & \overline{\psi_j}\delta(\omega) + i \sqrt{\frac{\pi\omega_m}{2}}\sum_{n=0}^N \frac{\delta(\omega-\frac{(2n+1)\omega_{m}}{2})}{\omega}  \label{eq:PI}\\
     \tilde{V}_{Cj}(\omega) = &  i(R_{Hj}I_0-2A_{2\omega_s}) \delta(\omega-\omega_s)+...  \nonumber \\ & +iA_{2\omega_s} \delta(\omega-2\omega_s) \label{eq:vc}\\
      \tilde{V}_Q(\omega) = & \sum_{j=0}^1 \tilde{\Pi}_j(\omega) * \tilde{V}_{Cj}(\omega) \label{eq:conv}
\end{align}
\par
By evaluating equation (\ref{eq:conv}) at the primary peak, we get equation (\ref{eq:primary}). From this, we expect resonant peaks in $\tilde{V}_Q(\omega_s)$ when $\omega_s=(2n+1)\omega_m/4\pi$. This frequency mixing is characteristic of the application of projective measurements, we do not expect to observe this in the classical case.

\begin{align}
\begin{split}
    &\tilde{V}_Q(\omega_s) =  \sum_{j=0}^1 (R_{Hj}I_0-A_{2\omega_s}) ...  \\ & \left( i\overline{\psi_j} - \sqrt{\frac{\pi\omega_m}{2}} \sum_{n=0}^N  \frac{1}{\omega_s} \delta(\omega_s- \frac{(2n+1)\omega_m}{4\pi})\right)  \label{eq:primary}
\end{split}
\end{align}

\section{Simulations}
\quad We use powers of two for many variables, since having $2^k$ data points is most efficient for calculations using the fast Fourier transform. We find appropriate convergence with a step size of $dt=2^{-14}s$, and chose a measurement frequency of 16Hz and a simulation time of T=32s. To look for periodicity, we collect data up to the $10^{th}$ harmonic of $\omega_m$. Each trial consists of 160 simulations of the device, 1 Hz$<\omega_s<160$ Hz. The maximum $\omega_s$ is within realistic bounds for timescales of ion diffusion \cite{https://doi.org/10.1002/adfm.201600680}. We use reported memristance parameters from ref. \cite{Stewart2004}: $R_{H0}=20k\Omega$,  $R_{L0}=2k\Omega$, $R_{H1}=2.1k\Omega$ $R_{L1}=250\Omega$. Ref. \cite{Stewart2004} also motivated the choice of $I_0$, since their device is stable for $V>$0.5V, we set $I_0=0.5/R_{H0}=25\times 10^{-6}$A.
\par
At each time-step $1<n<Tdt$, we calculate $V_n$, $\varphi_n$, $I_n$, $q_n$, and the wave function $\ket{\psi_n}$. These are calculated via the following, where all parameters are initialised at 0 (including $\ket{\psi}=\ket{0}$). Solutions to equations (\ref{eq:simphi}) and (\ref{eq:simpsi}) were computed using the 4$^{th}$ order Runge-Kutta method:
\begin{align}
    t_n=&t_{n-1}+dt \label{eq:simt}\\
    I_n=&I_0 \sin(\omega_s t_n)\label{eq:simi}\\
    q_n=&\frac{I_0}{\omega_s}(1-\cos(\omega_s t))\label{eq:simq}\\
    V_n=&I_n R_{Hj}(1-\frac{\mu_v}{D}R_{Lj}q_n)   \label{eq:simv}\\
    \frac{d\varphi_n}{dt}=&V_n \label{eq:simphi}\\
    \frac{d\psi_n}{dt}=&\frac{1}{i\hbar}H \ket{\psi_{n-1}}\label{eq:simpsi}
\end{align}

\section{Results\label{sec:res}}
\begin{figure}[h]
\centering
\includegraphics[width=\linewidth]{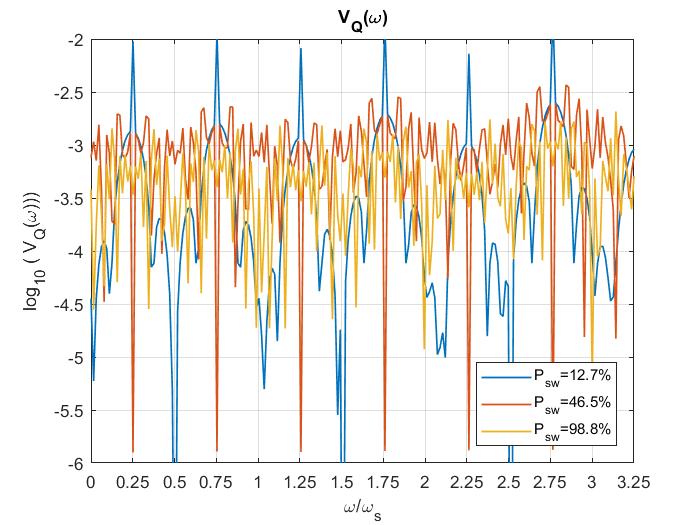}
\caption{Emergence of periodic spikes in the spectra as the probability of switching ($P_{sw}$) increases for $\omega_s$=100Hz. When $P_{sw}=0$, there is no real part to the spectrum. For low switching probabilities, the real part mimics the imaginary part, with a reduced amplitude. We see the emergence of the characteristic peaks/troughs at mid-range probabilities. At high $P_{sw}$ we see all resonances, the odd set having a higher amplitude than the even set.}
\label{fig:psw}
\end{figure}

\quad First, we look for the emergence of signal mixing in $\tilde{V}_Q(\omega)$ as $P_{sw}$ increases. This demonstrates that $P_{sw}$ is a mediator between the quantum and classical cases. In fig. \ref{fig:psw}, we see characteristic signal mixing of $\omega_m$ and $\omega_s$ for mid and high values of $P_{sw}$ \cite{Savelev2004}. The $P_{sw}=98.8\%$ data set is close to the switching function in fig. \ref{fig:squarewaves}a, the $P_{sw}=46.5\%$ data set has a similar to fig. \ref{fig:squarewaves}b. We do not illustrate the $P_{sw}=0$ example since there is no real component to the spectra in this case (deducible from equation (\ref{eq:classical})). 
\par
\begin{figure}[h]
    \centering
    \includegraphics[width=\linewidth]{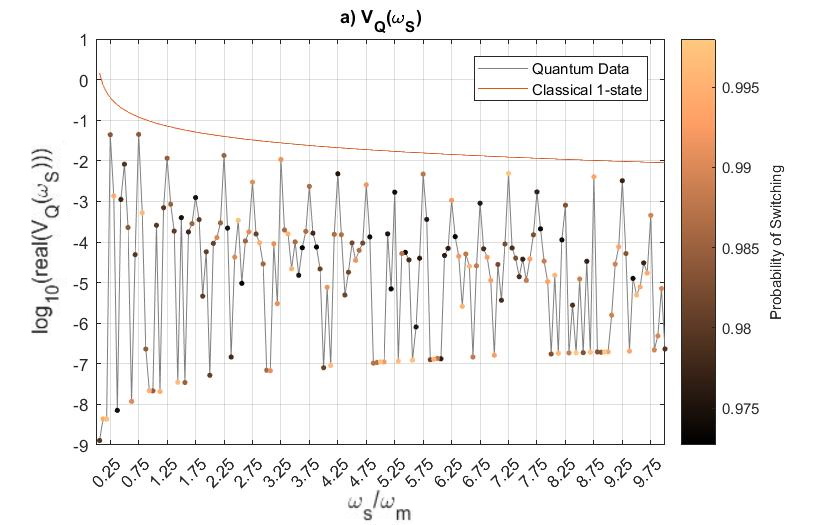}
    \includegraphics[width=\linewidth]{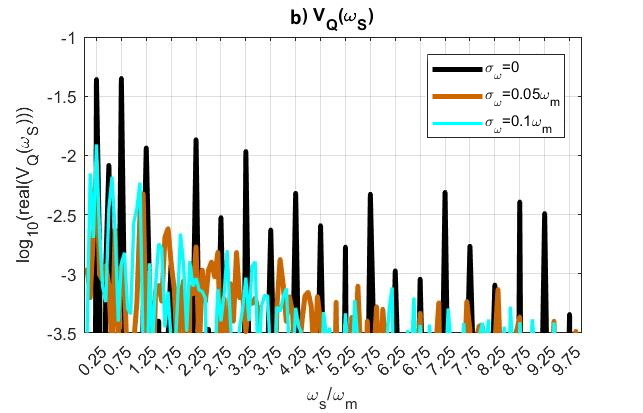}
    \caption{Real component of the primary peak as $\omega_s$ is varied. We see the signatures of the measurement process as predicted, peaks in the spectra at the key frequencies of $\omega_s=\frac{2n+1}{4\pi}\omega_m$. These are in addition to the peak at the square wave frequency, $\omega_s=\omega_m/2\pi$. \ref{fig:sweep}a shows the clear differences between the classical and quantum cases. \ref{fig:sweep}b demonstrates this method can be used in noisy systems, we still see peaks at many of the characteristic frequencies when there is a 10\% standard deviation in the measurement frequency.
    }\label{fig:sweep}
\end{figure}

Fig. \ref{fig:sweep} demonstrates how the real amplitude of the primary peak varies with the source frequency. In \ref{fig:sweep}a we compare the classical data (no peaks) with the quantum data (peaks at odd multiples of $\omega_m/4\pi$, see equation (\ref{eq:primary})) seeing a clear difference in behaviour. The average probability of switching $P_{sw}$ across this trial was 98.7$\pm$0.9\%. $P_{sw}$ is high enough for our model $\Pi_j$ function to be a suitable approximation. As such, we can confidently conclude that the presence of resonance at the predicted points is indicative of the application of regular projective measurements.
\par
Finally, we looked at cases with varying statistics of measurement events, to test the robustness of this methodology. We simulate trials where the standard deviations in $\omega_m$ is 5\% and 10\%. I.e. the measurement frequencies are normally distributed around $\omega_m$ by 0.05$\omega_m$ and 0.1$\omega_m$. $P_{sw}$ varies between each $\sigma_\omega$ trial: across the 5\% trials, $P_{sw}=0.889\pm0.014$; in the 10\% trials, $P_{sw}=0.810\pm0.017$. 
Fig. \ref{fig:sweep}b demonstrates the differences in peak heights and widths as expected. The higher $\sigma_\omega$, the larger the deviation from the idealised case, especially at high frequencies. In the 5\% trial, the first 7 peaks in the data set have similar positions to the $\sigma_\omega=0$ trial. In the 10\% trial, we see analogues for the first four peaks. In both cases, we clearly see the effects of signal mixing near the predicted points, showing our method is robust.

\section{Conclusions}
We have simulated systems to demonstrate our method to verify the presence of quantum coherent memristance states. Our model would fit several experimental systems, not limited to a van der Waals heterostructure \cite{VDW} or a singularly charged pair of quantum dots, with two memristive filaments per localised electron state. 
\par
We have demonstrated how the probability of switching state upon measurement controls the amount of quantumness we see in the voltage spectra. The higher the probability of switching, the more obvious the frequency mixing in real($V_q(\omega))$. \par 
We conduct frequency sweeps, tracking the real component of the amplitude of the primary spectral peak (fig. \ref{fig:sweep}a). This alternate method to find evidence of the periodic wave function collapse would be particularly useful in noisy systems, where the signal in the individual spectra may be shrouded. We see clear differences in behaviour in the classical and quantum schemes. Additionally, we carry out the same process for noisy systems, to test the robustness of the methodology. We still find evidence of frequency mixing even in the case where the measurement frequency varies by 10\%. This method is stable enough that the characteristic resonance patterns are still evident across all trials we have conducted.

\bibliography{mainbib}

\begin{thebibliography}{15}%
\makeatletter
\providecommand \@ifxundefined [1]{%
 \@ifx{#1\undefined}
}%
\providecommand \@ifnum [1]{%
 \ifnum #1\expandafter \@firstoftwo
 \else \expandafter \@secondoftwo
 \fi
}%
\providecommand \@ifx [1]{%
 \ifx #1\expandafter \@firstoftwo
 \else \expandafter \@secondoftwo
 \fi
}%
\providecommand \natexlab [1]{#1}%
\providecommand \enquote  [1]{``#1''}%
\providecommand \bibnamefont  [1]{#1}%
\providecommand \bibfnamefont [1]{#1}%
\providecommand \citenamefont [1]{#1}%
\providecommand \href@noop [0]{\@secondoftwo}%
\providecommand \href [0]{\begingroup \@sanitize@url \@href}%
\providecommand \@href[1]{\@@startlink{#1}\@@href}%
\providecommand \@@href[1]{\endgroup#1\@@endlink}%
\providecommand \@sanitize@url [0]{\catcode `\\12\catcode `\$12\catcode
  `\&12\catcode `\#12\catcode `\^12\catcode `\_12\catcode `\%12\relax}%
\providecommand \@@startlink[1]{}%
\providecommand \@@endlink[0]{}%
\providecommand \url  [0]{\begingroup\@sanitize@url \@url }%
\providecommand \@url [1]{\endgroup\@href {#1}{\urlprefix }}%
\providecommand \urlprefix  [0]{URL }%
\providecommand \Eprint [0]{\href }%
\providecommand \doibase [0]{http://dx.doi.org/}%
\providecommand \selectlanguage [0]{\@gobble}%
\providecommand \bibinfo  [0]{\@secondoftwo}%
\providecommand \bibfield  [0]{\@secondoftwo}%
\providecommand \translation [1]{[#1]}%
\providecommand \BibitemOpen [0]{}%
\providecommand \bibitemStop [0]{}%
\providecommand \bibitemNoStop [0]{.\EOS\space}%
\providecommand \EOS [0]{\spacefactor3000\relax}%
\providecommand \BibitemShut  [1]{\csname bibitem#1\endcsname}%
\let\auto@bib@innerbib\@empty
\bibitem [{\citenamefont {Chua}(1971)}]{Chua1971}%
  \BibitemOpen
  \bibfield  {author} {\bibinfo {author} {\bibfnamefont {L.}~\bibnamefont
  {Chua}},\ }\href {\doibase 10.1109/TCT.1971.1083337.} {\bibfield  {journal}
  {\bibinfo  {journal} {IEEE Transactions of Circuit Theory}\ }\textbf
  {\bibinfo {volume} {18}} (\bibinfo {year} {1971}),\
  10.1109/TCT.1971.1083337.}\BibitemShut {Stop}%
\bibitem [{\citenamefont {Chua}(2019)}]{Chua2019}%
  \BibitemOpen
  \bibfield  {author} {\bibinfo {author} {\bibfnamefont {L.}~\bibnamefont
  {Chua}},\ }\href {\doibase 10.1007/978-3-319-76375-0\_1} {\bibfield
  {journal} {\bibinfo  {journal} {Handbook of Memristor Networks}\ } (\bibinfo
  {year} {2019}),\ 10.1007/978-3-319-76375-0\_1}\BibitemShut {NoStop}%
\bibitem [{\citenamefont {Zagoskin}(2011)}]{zago11}%
  \BibitemOpen
  \bibfield  {author} {\bibinfo {author} {\bibfnamefont {A.~M.}\ \bibnamefont
  {Zagoskin}},\ }\href@noop {} {\emph {\bibinfo {title} {Quantum engineering:
  theory and design of quantum coherent structures}}}\ (\bibinfo  {publisher}
  {Cambridge University Press},\ \bibinfo {year} {2011})\BibitemShut {NoStop}%
\bibitem [{\citenamefont {Yi}\ \emph {et~al.}(2016)\citenamefont {Yi},
  \citenamefont {Savel'ev}, \citenamefont {Medeiros-Ribeiro},\ and\
  \citenamefont {et~al}}]{Yi2016}%
  \BibitemOpen
  \bibfield  {author} {\bibinfo {author} {\bibfnamefont {W.}~\bibnamefont
  {Yi}}, \bibinfo {author} {\bibfnamefont {S.}~\bibnamefont {Savel'ev}},
  \bibinfo {author} {\bibfnamefont {G.}~\bibnamefont {Medeiros-Ribeiro}}, \
  and\ \bibinfo {author} {\bibnamefont {et~al}},\ }\href {\doibase
  10.1038/ncomms11142} {\bibfield  {journal} {\bibinfo  {journal} {Nature
  Comms}\ } (\bibinfo {year} {2016}),\ 10.1038/ncomms11142}\BibitemShut
  {NoStop}%
\bibitem [{\citenamefont {Strukov}\ \emph {et~al.}(2008)\citenamefont
  {Strukov}, \citenamefont {Snider}, \citenamefont {Stewart},\ and\
  \citenamefont {et~al.}}]{Found2008}%
  \BibitemOpen
  \bibfield  {author} {\bibinfo {author} {\bibfnamefont {D.}~\bibnamefont
  {Strukov}}, \bibinfo {author} {\bibfnamefont {G.}~\bibnamefont {Snider}},
  \bibinfo {author} {\bibfnamefont {D.}~\bibnamefont {Stewart}}, \ and\
  \bibinfo {author} {\bibnamefont {et~al.}},\ }\href {\doibase
  10.1038/nature06932} {\bibfield  {journal} {\bibinfo  {journal} {Nature}\ }
  (\bibinfo {year} {2008}),\ 10.1038/nature06932}\BibitemShut {NoStop}%
\bibitem [{\citenamefont {Wang}\ and\ \citenamefont {et~al}(2017)}]{Wang2017}%
  \BibitemOpen
  \bibfield  {author} {\bibinfo {author} {\bibfnamefont {Z.}~\bibnamefont
  {Wang}}\ and\ \bibinfo {author} {\bibnamefont {et~al}},\ }\href {\doibase
  10.1038/nmat4756} {\bibfield  {journal} {\bibinfo  {journal} {Nature Mat}\ }
  (\bibinfo {year} {2017}),\ 10.1038/nmat4756}\BibitemShut {NoStop}%
\bibitem [{\citenamefont {Prezioso}\ \emph {et~al.}(2016)\citenamefont
  {Prezioso}, \citenamefont {Merrikh~Bayat}, \citenamefont {Hoskins},\ and\
  \citenamefont {et~al.}}]{Prezioso2016}%
  \BibitemOpen
  \bibfield  {author} {\bibinfo {author} {\bibfnamefont {M.}~\bibnamefont
  {Prezioso}}, \bibinfo {author} {\bibfnamefont {F.}~\bibnamefont
  {Merrikh~Bayat}}, \bibinfo {author} {\bibfnamefont {B.}~\bibnamefont
  {Hoskins}}, \ and\ \bibinfo {author} {\bibnamefont {et~al.}},\ }\href
  {\doibase 10.1038/srep21331} {\bibfield  {journal} {\bibinfo  {journal}
  {Scientific Reports}\ } (\bibinfo {year} {2016}),\
  10.1038/srep21331}\BibitemShut {NoStop}%
\bibitem [{\citenamefont {Jiang}\ and\ \citenamefont
  {et~al}(2017)}]{Jiang2017}%
  \BibitemOpen
  \bibfield  {author} {\bibinfo {author} {\bibfnamefont {H.}~\bibnamefont
  {Jiang}}\ and\ \bibinfo {author} {\bibnamefont {et~al}},\ }\href {\doibase
  10.1038/s41467-017-00869-x} {\bibfield  {journal} {\bibinfo  {journal}
  {Nature Comms}\ } (\bibinfo {year} {2017}),\
  10.1038/s41467-017-00869-x}\BibitemShut {NoStop}%
\bibitem [{\citenamefont {Ushakov}\ \emph {et~al.}(2021)\citenamefont
  {Ushakov}, \citenamefont {Balanov},\ and\ \citenamefont
  {Savel’ev}}]{USHAKOV2021110803}%
  \BibitemOpen
  \bibfield  {author} {\bibinfo {author} {\bibfnamefont {Y.}~\bibnamefont
  {Ushakov}}, \bibinfo {author} {\bibfnamefont {A.}~\bibnamefont {Balanov}}, \
  and\ \bibinfo {author} {\bibfnamefont {S.}~\bibnamefont {Savel’ev}},\
  }\href {\doibase https://doi.org/10.1016/j.chaos.2021.110803} {\bibfield
  {journal} {\bibinfo  {journal} {Chaos, Solitons \& Fractals}\ } (\bibinfo
  {year} {2021}),\ https://doi.org/10.1016/j.chaos.2021.110803}\BibitemShut
  {NoStop}%
\bibitem [{\citenamefont {Ushakov}\ and\ \citenamefont
  {et~al}(2021)}]{USHAKOV2021110997}%
  \BibitemOpen
  \bibfield  {author} {\bibinfo {author} {\bibfnamefont {Y.}~\bibnamefont
  {Ushakov}}\ and\ \bibinfo {author} {\bibnamefont {et~al}},\ }\href {\doibase
  https://doi.org/10.1016/j.chaos.2021.110997} {\bibfield  {journal} {\bibinfo
  {journal} {Chaos, Solitons \& Fractals}\ } (\bibinfo {year} {2021}),\
  https://doi.org/10.1016/j.chaos.2021.110997}\BibitemShut {NoStop}%
\bibitem [{\citenamefont {Choi}\ and\ \citenamefont
  {et~al}(2016)}]{https://doi.org/10.1002/adfm.201600680}%
  \BibitemOpen
  \bibfield  {author} {\bibinfo {author} {\bibfnamefont {B.}~\bibnamefont
  {Choi}}\ and\ \bibinfo {author} {\bibnamefont {et~al}},\ }\href {\doibase
  10.1002/adfm.201600680} {\bibfield  {journal} {\bibinfo  {journal} {Advanced
  Functional Materials}\ } (\bibinfo {year} {2016}),\
  10.1002/adfm.201600680}\BibitemShut {NoStop}%
\bibitem [{\citenamefont {Sun}\ and\ \citenamefont {et~al}(2019)}]{Sun2019}%
  \BibitemOpen
  \bibfield  {author} {\bibinfo {author} {\bibfnamefont {W.}~\bibnamefont
  {Sun}}\ and\ \bibinfo {author} {\bibnamefont {et~al}},\ }\href {\doibase
  10.1038/s41467-019-11411-6} {\bibfield  {journal} {\bibinfo  {journal}
  {Nature Comms}\ } (\bibinfo {year} {2019}),\
  10.1038/s41467-019-11411-6}\BibitemShut {NoStop}%
\bibitem [{\citenamefont {Lucatto}\ and\ \citenamefont {et~al}(2019)}]{VDW}%
  \BibitemOpen
  \bibfield  {author} {\bibinfo {author} {\bibfnamefont {B.}~\bibnamefont
  {Lucatto}}\ and\ \bibinfo {author} {\bibnamefont {et~al}},\ }\href {\doibase
  10.1103/PhysRevB.100.121406} {\bibfield  {journal} {\bibinfo  {journal}
  {Phys. Rev. B}\ } (\bibinfo {year} {2019}),\
  10.1103/PhysRevB.100.121406}\BibitemShut {NoStop}%
\bibitem [{\citenamefont {Stewart}\ and\ \citenamefont
  {et~al}(2004)}]{Stewart2004}%
  \BibitemOpen
  \bibfield  {author} {\bibinfo {author} {\bibfnamefont {D.}~\bibnamefont
  {Stewart}}\ and\ \bibinfo {author} {\bibnamefont {et~al}},\ }\href {\doibase
  10.1021/nl034795u} {\bibfield  {journal} {\bibinfo  {journal} {Nano Letters}\
  } (\bibinfo {year} {2004}),\ 10.1021/nl034795u}\BibitemShut {NoStop}%
\bibitem [{\citenamefont {Savel'ev}\ and\ \citenamefont
  {et~al}(2004)}]{Savelev2004}%
  \BibitemOpen
  \bibfield  {author} {\bibinfo {author} {\bibfnamefont {S.}~\bibnamefont
  {Savel'ev}}\ and\ \bibinfo {author} {\bibnamefont {et~al}},\ }\href {\doibase
  10.1140/epjb/e2004-00208-8} {\bibfield  {journal} {\bibinfo  {journal}
  {European Physical Journal B}\ } (\bibinfo {year} {2004}),\
  10.1140/epjb/e2004-00208-8}\BibitemShut {NoStop}%
\end{thebibliography}%

\end{document}